# The orbital effect on the anomalous magnetism and evolution in La$_x$Y$_{1-x}$VO$_3$ (0 ≤ $x$ ≤ 0.2) single crystals

Y. Wan[1], J. Jiao[1], G. Lin[1], Q. Sun[2], G. Wang[1], J. Zhu[1], B. Zhao[1], Q. Ren[3,4], M. Zhang[1], M. Chen[1], R. Zhang[5], W. Tong[5], J. Weissenrieder[6], X. Yao[1,7] *, J. Ma[1,7] *

[1] Key Laboratory of Artificial Structures and Quantum Control, Shenyang National Laboratory for Materials Science, School of Physics and Astronomy, Shanghai Jiao Tong University, Shanghai 200240, China

[2] Department of Mechanical Engineering, University of California, Riverside, California 92521, USA

[3] Institute of High Energy Physics, Chinese Academy of Sciences (CAS), Beijing 100049, China

[4] Spallation Neutron Source Science Center, Dongguan 523803, China

[5] Anhui Province Key Laboratory of Condensed Matter Physics at Extreme Conditions, High Magnetic Field Laboratory, HFIPS, Chinese Academy of Sciences, Hefei 230031, China

[6] Materials and Nano Physics, School of Engineering Sciences, KTH Royal Institute of Technology, SE-100 44 Stockholm, Sweden

[7] Collaborative Innovation Center of Advanced Microstructures, Nanjing University, Nanjing 210093, China

* Author to whom correspondence should be addressed.

E-mail address: xyao@sjtu.edu.cn (X. Yao); jma3@sjtu.edu.cn (J. Ma)

**Abstract:** The orbital effect on the anomalous magnetism and evolution of low La-doping single crystals, La$_x$Y$_{1-x}$VO$_3$ ($x$ = 0, 0.1, and 0.2), has been investigated by applying the X-ray diffraction, specific heat, magnetization and Raman scattering techniques. The larger nearest-neighbor (NN) exchange interaction along c-axis stabilizes the fluctuant G-type orbital ordering (G-OO) which favors the exotic C-type antiferromagnetic order (C-AF). It is found that the NN exchange interaction in ab plane is anisotropy relating to the in plane magnetic anisotropy, which becomes smaller in high La-doped sample. Most interestingly, with increasing the La$^{3+}$ content the orbital fluctuation and hybridization are decreased which stabilizes the C-OO phase and destabilizes the G-OO phase. Meanwhile, the diamagnetism in the exotic C-AF phase becomes weak and the possible mechanism relates to the change of the competition between the single-ion magnetic anisotropy and the Dzyaloshinsky-Moriya (DM) interaction with increasing $x$. Finally, the strong spin-orbital coupling has been observed at temperature just above $T_N$ in $La_{0.2}Y_{0.8}VO_3$ and a short range spin-orbital correlation is suggested.

## 1. Introduction

Transition-metal oxides (TMOs) with perovskite-related structure ($ABO_3$), for which the interactions of electron, spin, orbital and lattice are complicated, could demonstrate a variety of intriguing physical properties such as high-$T_c$ superconductivity, Mott transition, and colossal magnetoresistance [1-6]. Due to its specific ion environment [7], the vanadate, $RVO_3$ (R=$Y^{3+}$, $La^{3+}$, etc.), is a typical platform to study the complicated spin-orbital interaction [8]: in the $VO_6$ octahedra, the degenerate $V^{3+}$ ($3d^2$) $t_{2g}$ orbitals split into a low-energy singlet with occupation of $d_{xy}$ orbital and a high-energy doublet with alternated occupation of $d_{xz}$ and $d_{yz}$ orbitals, respectively, which induce a Jahn-Teller (JT) distortion and anisotropic V-O bonds [9]. Figure 1 (c) present the alternating occupied $d_{xz}$ or $d_{yz}$ orbitals as G-type orbital ordering (G-OO) or C-type orbital ordering (C-OO) [10, 11]. From the spin-orbital-lattice coupling and the Goodenough-Kanamori-Anderson rule [12, 13], the G-OO induces the C-type antiferromagnetic ordering (C-AF), while the C-OO favors the G-type antiferromagnetic ordering (G-AF).

If the $R^{3+}$ ions are nonmagnetic $Y^{3+}$ and $La^{3+}$ ions, no extra magnetic environment interferes with the magnetism of $VO_3$ and only the lattice effect is demonstrated, hence the magnetism of $V^{3+}$ ions could be adjusted by the spin-orbital-lattice coupling with the different radius of $Y^{3+}$ and $La^{3+}$ ions [14]. For $YVO_3$, the G-OO state is accompanied by the change of the crystal structure from distorted orthorhombic Pnma to monoclinic $P2_1/c$ with temperature cooling across $T_{OO}$, followed by C-AF at $T_N$ [15]. With continuous cooling, the orbital/spin-ordering transform from C-AF/G-OO to G-AF/C-OO at $T_{CG}$, and the lattice structure was back to the space group of Pnma [16]. Although the C-AF/G-OO phase between $T_{CG}$ and $T_N$ is accompanied with the magnetization reversal in $YVO_3$ single crystals [10, 17-19], the physical mechanism to explain this anomalous magnetism is still under debate [20]. As $La^{3+}$ ion is doped on Y-site, the enlarged average ion radius of $R^{3+}$ not only affect the bond length of V-O and the bond angle of V-O-V, but also adjust the nearest-neighbor (NN) exchange energy of $V^{3+}$ ions and the magnetic anisotropy dramatically.

Yan, *et.al.* [14] have studied the system of $La_xY_{1-x}VO_3$ ($0 \leq x \leq 1$) and built up the magnetic phase diagram of composition versus temperature. As $0 \leq x < 0.2$, $T_{CG}$ goes up while $T_{OO}$ and $T_N$ decrease gradually; as $x > 0.2$, only the G-AF/C-OO phase is accompanied by the orthorhombic phase below $T_N$, while the C-AF/G-OO phase and the magnetization reversal feature disappeared [14]. Therefore, $La_{0.2}Y_{0.8}VO_3$ ($x$=0.2) is an important critical composition with the dramatically different magnetic proprieties and both $T_{CG}$ and $T_{OO}$ have been reported to disappear. However, Yano, *et.al.* [21] suggested that $T_{CG}$ and $T_{OO}$ could be very close and coexist in $La_{0.2}Y_{0.8}VO_3$. Therefore, a detailed investigation of temperature dependence of the magnetic properties and its lattice response of $La_{0.2}Y_{0.8}VO_3$ is needed, which will help us understand the anomalous magnetism and the related evolution in $La_xY_{1-x}VO_3$ single crystals with the lattice precisely adjust by $La^{3+}$ ions.

To investigate this issue, the La-doping single crystals, $La_xY_{1-x}VO_3$ ($x$ = 0, 0.1 and 0.2), were synthesized to precisely adjust the $V^{3+}$ orbital contribution. With the measurements of the X-ray diffraction, specific heat, magnetization and Raman scattering, the magnetic anisotropy presented that NN exchange interaction along *c*-axis stabilized the fluctuant G-OO and favored the exotic C-AF phase, while the anisotropy NN exchange interaction in *ab* plane becomes smaller in higher doped sample. Moreover, the shrinking of the magnetization reversal and C-AF/G-OO phases were observed as a function of *x*, and the interplay of the single-ion magnetic anisotropy and the Dzyaloshinsky-Moriya (DM) interaction was revealed as a magnetization inflection temperature point (~91K) for $La_{0.1}Y_{0.9}VO_3$ under a modest applied field.

## 2. Experimental Section

A two-mirror vertical image furnace (Quantum Design Corporation) was applied to synthesize $La_xY_{1-x}VO_3$ ($x$=0, 0.1, 0.2) single crystals by melting-grown from polycrystalline rods, which were mixed by $Y_2O_3$, $V_2O_3$ and $La_2O_3$ (Alfa Aesar, 99.99%) powders in stoichiometric ratio and heated at 1450°C for 15h under argon atmosphere. The phase purity of the poly-crystalline sample was checked by powder XRD after the heating process. During the crystal growth, a high purity argon pressure of 5 bar was used to prevent the as-grown crystals to crack and oxidize. The growth rate was 3~5 mm/h with downward direction, and the rotation of the up and down rods were 15 rpm in opposite directions. The small pieces of the single-crystals were grounded to check the phases by powder XRD and the orthorhombic Pnma phase was confirmed. The x-ray Laue diffractometer was applied to determine the orientation of the single-crystal for the investigation of magnetization, thermal and Raman optical properties.

The magnetization and specific heat measurements were carried out with a Physical Properties Measurement System (PPMS, Quantum Design) in the temperature interval 3~300 K. The Polarized Raman-scattering experiments were performed with using a Raman Microscope (Horiba JY T64000), and the low-temperature was obtained by a Janis ST-500 microscopy cryostat for the temperature control from 5 to 300 K. A 532 nm light from a He-Ne laser was focused onto a 0.05-mm-diameter spot on the sample surface. The scattering light was detected by using a CCD with a spectral resolution of 1 cm$^{-1}$. The spectra were measured in a polarization configuration of $a(bb)a$, where the notation of $a$ and $b$ represent the polarization and direction of incident or scattering light along the $a$- and $b$- axis of the single crystal, respectively.

## 3. Results

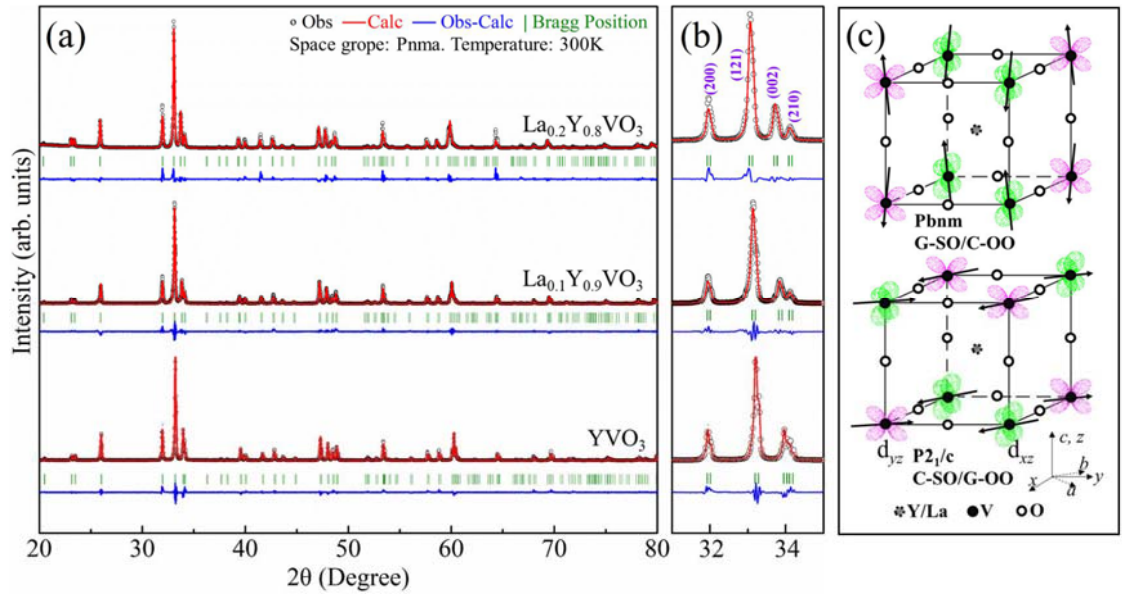

Fig.1. (a) Room temperature X-ray powder diffraction of La$_x$Y$_{1-x}$VO$_3$ (x=0~0.2). (b) The enlarged figure in the 2θ range of 31°~35°. (c) Schematic structures of the G-type antiferromagnetic ordering (G-AF) and C-type orbital ordering (C-OO), the C-AF and G-OO in the perovskite-type YVO$_3$.

The XRD patterns of La$_x$Y$_{1-x}$VO$_3$ (x: 0~0.2) were shown in Fig.1 (a) and (b). The lattice structure was simulated by the orthorhombic Pnma space group with the FullProf program, and no impurities were observed. When increasing the concentration of La$^{3+}$ ions, both the peak (121) shifting to lower angles and the separation of the (002) and (210) peaks suggest a lattice expansion with the substitution for Y$^{3+}$ ions by the larger La$^{3+}$ ions in host lattice (Y$^{3+}$:1.075A, La$^{3+}$:1.216A) [22]. Furthermore, both the NN V-V bond length along the $c$-axis (VV$_c$) and in $ab$ plane (VV$_{ab}$) enlarge with increasing $x$ from the XRD patterns, Table 1.

Table 1. The NN V-V band length obtained from the XRD refinement data of $La_xY_{1-x}VO_3$ ($0 \leq x \leq 0.2$)

|  | $YVO_3$ | $La_{0.1}Y_{0.9}VO_3$ | $La_{0.2}Y_{0.8}VO_3$ |
|---|---|---|---|
| $VV_{ab}$ (Å) | 3.848 | 3.853 | 3.861 |
| $VV_c$ (Å) | 3.787 | 3.795 | 3.807 |

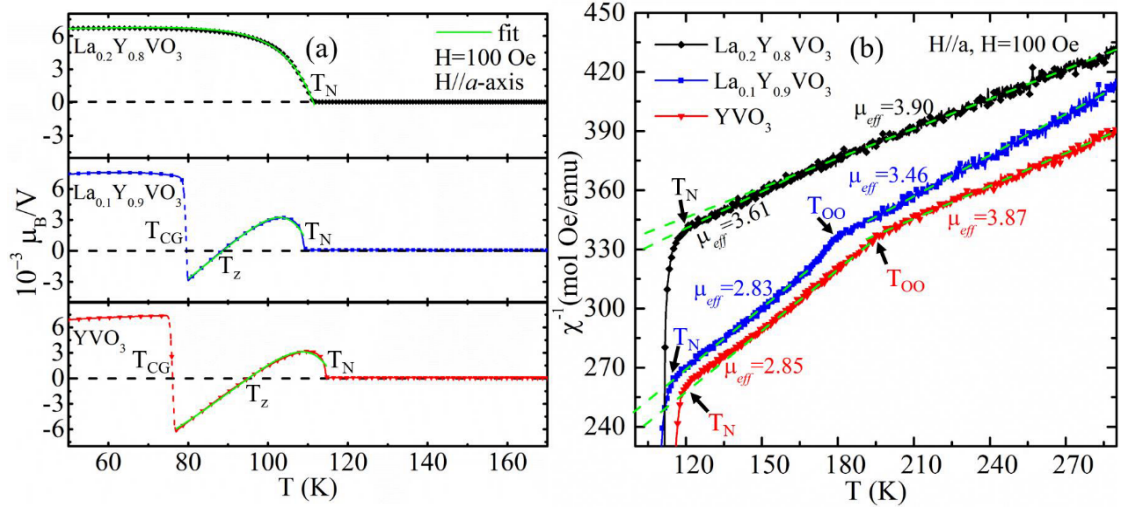

Fig. 2. Temperature dependence of the magnetization (a) and inverse susceptibility (b) with the applied field (H=100 Oe) along *a*-axis for the single crystals of $La_xY_{1-x}VO_3$ (x=0~0.2), respectively. The green lines are calculated using the equation with the corresponding fit functions.

The temperature dependence of magnetization along *a*-axis of $La_xY_{1-x}VO_3$ ($0 \leq x \leq 0.2$) are shown in Fig. 2 (a). The magnetization curves show the dramatic jump from negative to positive value at $T_{CG}$ of $YVO_3$ and $La_{0.1}Y_{0.9}VO_3$. As *x* increase from 0 to 0.1, the intensity of the jump reduces, $T_{CG}$ increases and $T_N$ decreases. In $La_{0.2}Y_{0.8}VO_3$, $T_{CG} \sim T_N$ is replaced by $T_N$, which indicates the disappearance of the C-AF/G-OO phase and the magnetization reversal. These observations suggest that the substitution for $Y^{3+}$ by $La^{3+}$ ions can enhance the stability of G-AF/C-OO phase and shrink the C-AF/G-OO phase.

The inverse magnetic susceptibilities of $La_xY_{1-x}VO_3$ ($0 \leq x \leq 0.2$) along the *a*-axis above $T_N$ are plotted with the fitting by the Curie-Weiss law $\chi=C/(T+\theta)$, Fig. 2(b), it is unambiguous that $T_{OO}$ drops with *x* increasing. Above $T_{OO}$, the effective moments ($\mu_{eff}= g\sqrt{S(S+1)}$) are 3.46 $\mu_B$ and 3.87 $\mu_B$, which signals a strong spin-orbital coupling in $La_{0.1}Y_{0.9}VO_3$ and $YVO_3$, respectively. Between $T_N$ and $T_{OO}$, the effective moments decrease to 2.83 $\mu_B$ and 2.85 $\mu_B$ in $La_{0.1}Y_{0.9}VO_3$ and $YVO_3$, respectively, which are approximate to 2.828 $\mu_B$ (g=2) for the $V^{3+}$ ions (S=1). Therefore, the orbital is quenched and there is no spin-orbital coupling in both two compounds. For $La_{0.2}Y_{0.8}VO_3$, an inflection point around 150 K is observed though it is not obvious as illustrated in Fig. 2(b). The effective moment is 3.90 $\mu_B$ above 150 K and decreases to 3.61 $\mu_B$ between $T_N$ and 150 K, which suggests a strong spin-orbital coupling unlike with $La_{0.1}Y_{0.9}VO_3$ and $YVO_3$ [14, 21].

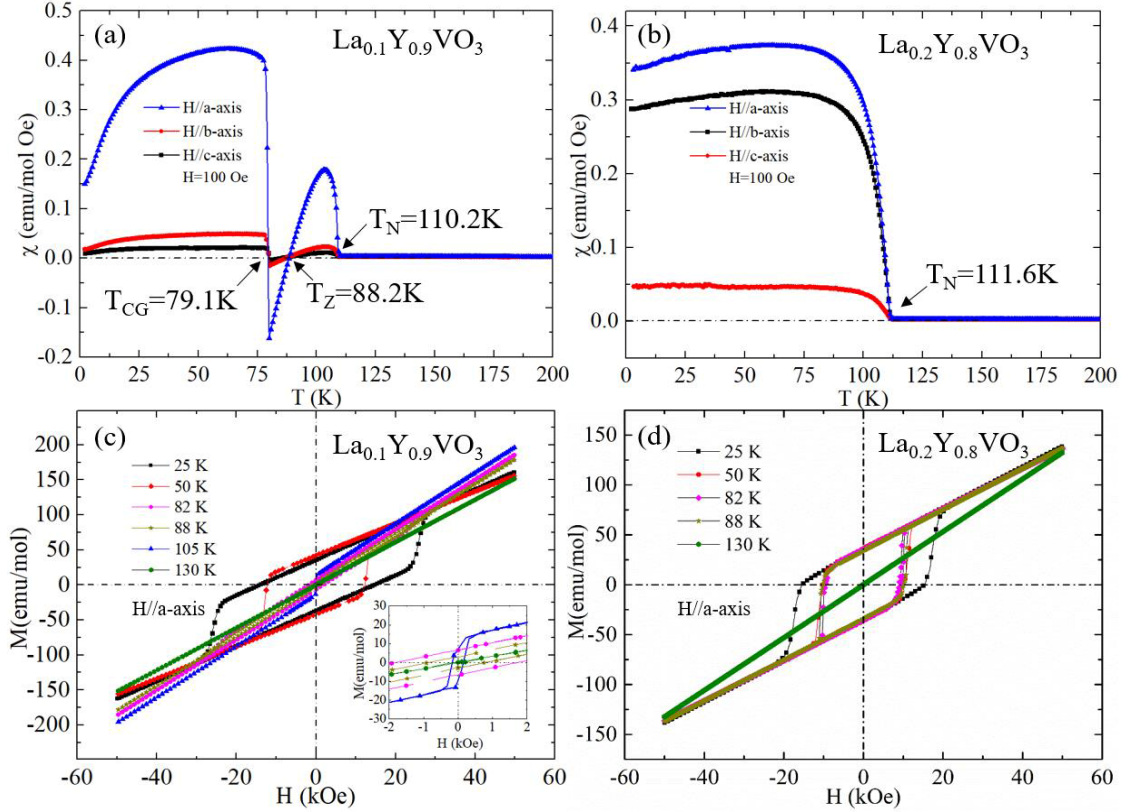

Fig. 3. Temperature vs. magnetization in an applied magnetic field of 100 Oe along *a*, *b*, and *c* -axis of $La_{0.1}Y_{0.9}VO_3$ (a) and $La_{0.2}Y_{0.8}VO_3$ (b) single crystals. Applied field dependence of magnetization along *a*-axis of $La_{0.1}Y_{0.9}VO_3$ (*c*) and $La_{0.2}Y_{0.8}VO_3$ (*d*) single crystals at different temperatures.

The temperature dependence of magnetization along the *a*-, *b*-, and *c*-axes as field-cooling (FC) for $La_{0.1}Y_{0.9}VO_3$ and $La_{0.2}Y_{0.8}VO_3$ single crystals were measured at 100 Oe, respectively, Fig. 3. For $La_{0.1}Y_{0.9}VO_3$, the FC magnetization along three axes show the sharp transition around $T_{CG}$=79.1 K and the anisotropic "diamagnet" at $T_Z$. For $La_{0.2}Y_{0.8}VO_3$, the antiferromagnetic transition is $T_N$=111.6 K. Meanwhile, the absolute magnetizations for both crystals clearly demonstrate the magnetic anisotropy with the largest value along the *a*-axis.

Figure 3 (c) displays the field-dependence of magnetization along the *a*-axis of $La_{0.1}Y_{0.9}VO_3$, and the positive differential susceptibility dM/dH indicates that the negative magnetization between $T_{CG}$ and $T_Z$ is not from a conventional diamagnetism. Since two groups of high-field dM/dH slopes at T=25K/50K and T=82K/88K/105K are same, the sublattice magnetic moments should be perpendicular to the *a*-axis at

$T_{CG}$=79.1 K. At 130 K > $T_N$, the high-field slopes dM/dH decreases and the related transition from the spin ordering to disordering. Moreover, as shown in Fig. 3 (c) it is clear that the remanent (net) magnetic moment and the coercive force between $T_{CG}$ and $T_N$ are much smaller than that below $T_{CG}$, associating with the strongly reduced magnetization with temperature just beyond $T_{CG}$, Fig. 3 (a). This observation relates to the different magnetic moment canting angles along the *a*-axis below and above $T_{CG}$. What is more, as shown in Fig. 3(c) and (d), M does not saturate even up to 5 T, which may attribute to the Van Vleck contribution.

For $La_{0.2}Y_{0.8}VO_3$, the high-field slopes dM/dH at T=25, 50, 82 and 88K<$T_N$, Fig. 3 (d), and that the sublattice magnetic moments are perpendicular to the *a*-axis as $La_{0.1}Y_{0.9}VO_3$. At 130K>$T_N$, the high-field slope is different to the low temperature slope and a first ordering transition is illustrated at $T_N$. Furthermore, the hysteresis loops show a coercivity of about 10.9 kOe and a sharp knee of the loops at an $H_n$≈±9.2 kOe for 50, 82 and 88K <$T_N$. Nucleation of the reverse domains needs a negative field |H−$H_n$|>0, and the relatively square M-H loop with the sharp knee suggests that domain walls are quickly mobile in the fields H>$H_n$.

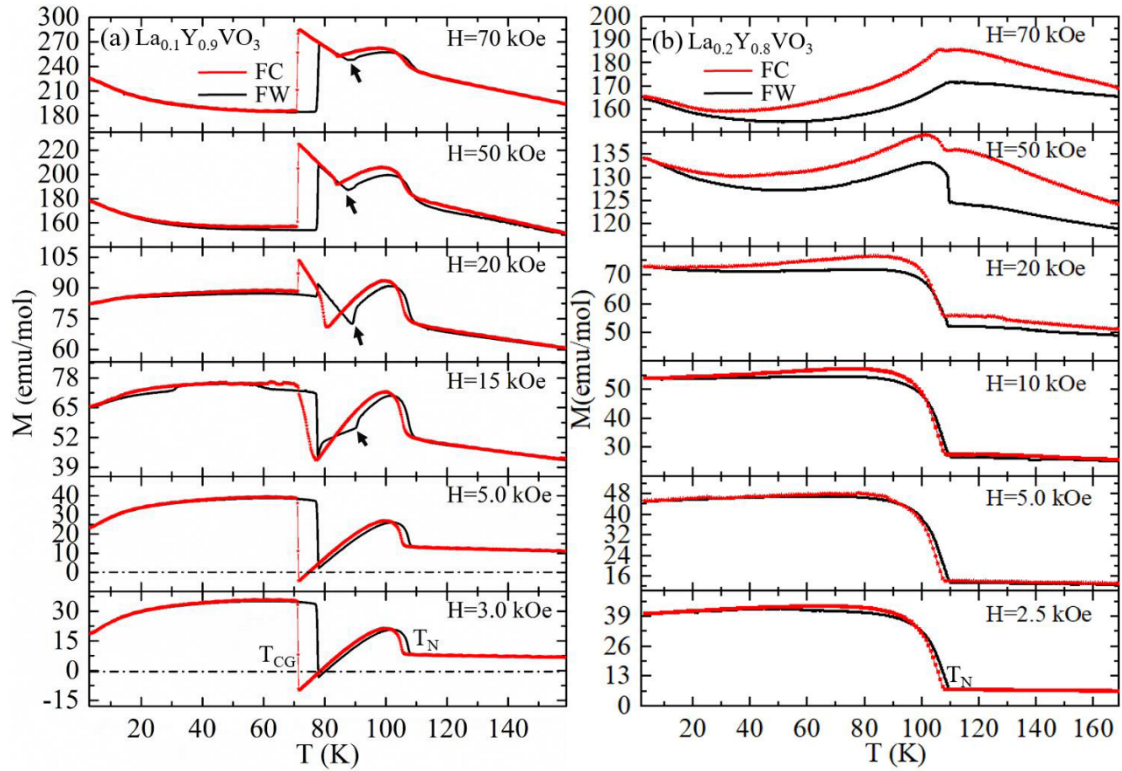

Fig. 4. Temperature dependence of the magnetization in different applied field along *a*-axis for the single crystals of $La_{0.1}Y_{0.9}VO_3$ (a) and $La_{0.2}Y_{0.8}VO_3$ (b). The data were measured in each field during field-cooling (FC) and then field-warming (FW).

Figure 4 illustrates the M~T curves along the *a*-axis with different applied fields in FC and FW for $La_{0.1}Y_{0.9}VO_3$ and $La_{0.2}Y_{0.8}VO_3$. The magnetization transitions with H = 3.0 kOe at $T_{CG}$ and $T_N$ demonstrate the thermal hysteresis for $La_{0.1}Y_{0.9}VO_3$ and a first-order transition is suggested. Below the two transition temperatures, the sharp increases of the magnetization indicate the ferromagnetic component below $T_{CG}$ and $T_N$. With increasing the applied field, the magnetization would not become negative at 5 kOe (FW), and the minimum magnetization remains at $T_{CG}$, Fig. 4(a). As the field enhanced to 15 kOe, a new inflection point (~91 K) appeared in the FW curve and the minimum magnetization point for the FC curve changed from 72K to 78 K. As the field increased to 20 kOe, the inflection point in the FW curve below $T_N$ instead the minimum magnetization point at $T_{CG}$. As the field increased continuously, the inflection point of FW and FC curves between $T_{CG}$ and $T_N$ increases monotonously.

However, for $YVO_3$ the inflection point appears in M~T curve with the applied field just larger than 5 kOe [23-25], which is smaller than the 15 kOe of $La_{0.1}Y_{0.9}VO_3$. This observation suggests a higher superexchange interaction between NN $V^{3+}$ ions along a-axis in $La_{0.1}Y_{0.9}VO_3$. And the formation of the inflection point in the modest field may relate with the reversed canting component along the a-axis of the spin moment caused by the DM interaction. $La_{0.2}Y_{0.8}VO_3$, presents a clear magnetic hysteresis just below $T_N$, but the degree is small than $La_{0.1}Y_{0.9}VO_3$, which relates with the thermal hysteresis existing in the latter at $T_{CG}$. If the field arises, the feature of M-T curve is gradually changing from ferromagnet to antiferromagnet and the noncollinear magnetic moments is observed, Fig. 4(b).

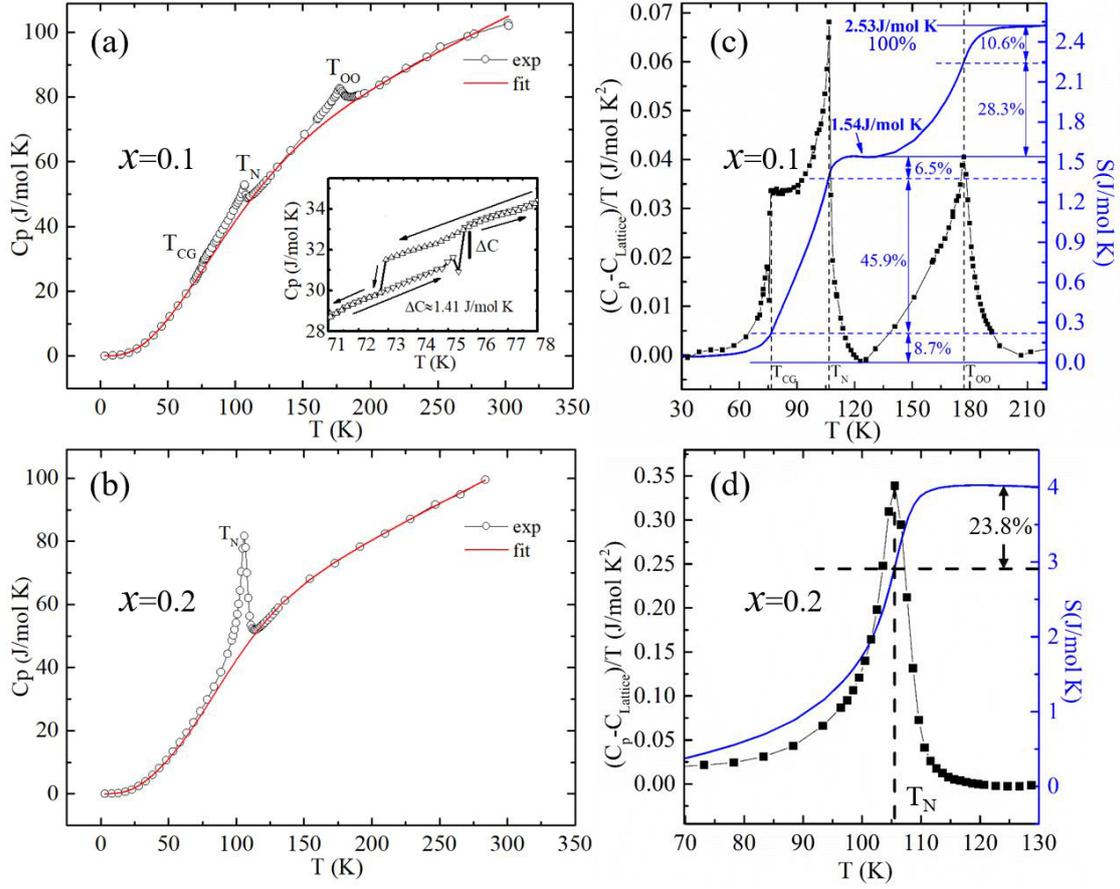

Fig. 5. Temperature dependence of specific heat ($C_p$) for $La_{0.1}Y_{0.9}VO_3$ (a) and $La_{0.2}Y_{0.8}VO_3$ (b) single crystals. The solid red lines in (a) and (b) illustrate the fit of lattice contribution ($C_{lattice}$) with the Debye and Einstein functions. Temperature dependence of ($C_p-C_{lattice}$)/T and the numerical integration of the entropy S after subtracting the lattice contribution for $La_{0.1}Y_{0.9}VO_3$ (c) and $La_{0.2}Y_{0.8}VO_3$ (d) single crystals.

Figure 5 (a) illustrates the specific-heat data for $La_{0.1}Y_{0.9}VO_3$ with three transition temperatures $T_{CG}$, $T_N$, and $T_{OO}$ unambiguously observed, and the structural/magnetic transition induces a dramatic increase in specific-heat of $\Delta C$=1.41 J/ (mol K) at $T_{CG} \approx$75.6 K, which is smaller than that of $YVO_3$ ($\Delta C$=3.53 J/ (mol K) [9]). For the crystal of $La_{0.2}Y_{0.8}VO_3$, there is only one transition temperature at $T_N$=105.8 K, Fig. 5 (b). Therefore, no structural transition is observed at $T_N$ for $La_{0.2}Y_{0.8}VO_3$.

To obtain the influence of the orbital and/or spin ordering on $La_xY_{1-x}VO_3$, the lattice contribution of the specific-heat data needed to be estimated and subtracted. The lattices contribution was fitted with the Debye and Einstein terms.

$$C_{lat}(T) = 9Nk_B\alpha_D (\frac{T}{\theta_D})^3 \int_0^{\frac{\theta_D}{T}} \frac{x^4 e^x}{(e^x-1)^2} dx + 3Nk_B\alpha_{E1} \frac{(\frac{\theta_{E1}}{T})^2 \exp(\frac{\theta_{E1}}{T})}{[\exp(\frac{\theta_{E1}}{T})-1]^2} +$$

$$3Nk_B\alpha_{E2} \frac{(\frac{\theta_{E2}}{T})^2 \exp(\frac{\theta_{E2}}{T})}{[\exp(\frac{\theta_{E2}}{T})-1]^2} + 3Nk_B\alpha_{E3} \frac{(\frac{\theta_{E3}}{T})^2 \exp(\frac{\theta_{E3}}{T})}{[\exp(\frac{\theta_{E3}}{T})-1]^2} \qquad (1)$$

where $\theta_D$ and $\theta_E$ are the Debye and Einstein temperatures, respectively. And $\alpha_D$ and $\alpha_E$

are the Debye and Einstein coefficients, respectively. The Debye function originates from the low-energy acoustic modes, while the Einstein function can estimate high-frequency optical modes.

The estimated curve for $C_p \sim T$ of $La_{0.1}Y_{0.9}VO_3$ is illustrated in Fig. 5(a), and the two lambda-shaped anomalies at $T_N \sim 106.8$ K and $T_{OO} \sim 177.5$ K. Figure 5(c) presents the temperature-dependence of the entropy changes and the entropy change of 1.16 J/(mole K) between $T_{CG}$ and $T_N$ are smaller than the expected value of $S_{mag}$=9.13 J/(mole K) ($S_{mag}$=Rln(2S+1), R=8.314) which was used to estimate the magnetic ordering entropy change for an isotropic spin system, indicating a spin-orbital coupling existing in this temperature range. The entropy change associated with the G-type orbital ordering was estimated to be 0.984 J/(mole K), which is smaller than that for $YVO_3$ [9]. Meanwhile, the G-type orbital ordering temperature decreases from $YVO_3$ to $La_{0.1}Y_{0.9}VO_3$, and La doping reduces the energy of orbital interaction. Although C-AF/G-OO has been formed, the entropy change exist in the whole range between $T_{CG}$ and $T_N$ with spin and/or orbital fluctuation, Fig. 5 (a) and (c).

For $La_{0.2}Y_{0.8}VO_3$, the entropy change is about 4.02 J/(mole K) around $T_N$ as shown in Fig. 5 (d), which is about twice the total entropy change (2.31 J/mole K) of $La_{0.1}Y_{0.9}VO_3$ at $T_{OO}$ and $T_N$, therefore, the extra effect exceeding a magnetic ordering was suggested at the $T_N$. Above $T_N$, a larger amount of entropy is released for $x$=0.20, which indicates that significant fluctuations related with either orbital or spin persist above $T_N$ for $x$=0.20. Short-range spin correlations stabilize orbital ordering [14], which would introduce regions of strong spin-orbit coupling above $T_N$. Meanwhile, the long-range orbital order was suppressed below $T_N$. With $x$ increasing to 0.5, the percentage of the entropy release above $T_N$ increases [14], which suggests the decreasing of the spin-orbit interaction and the enhancement of C-OO.

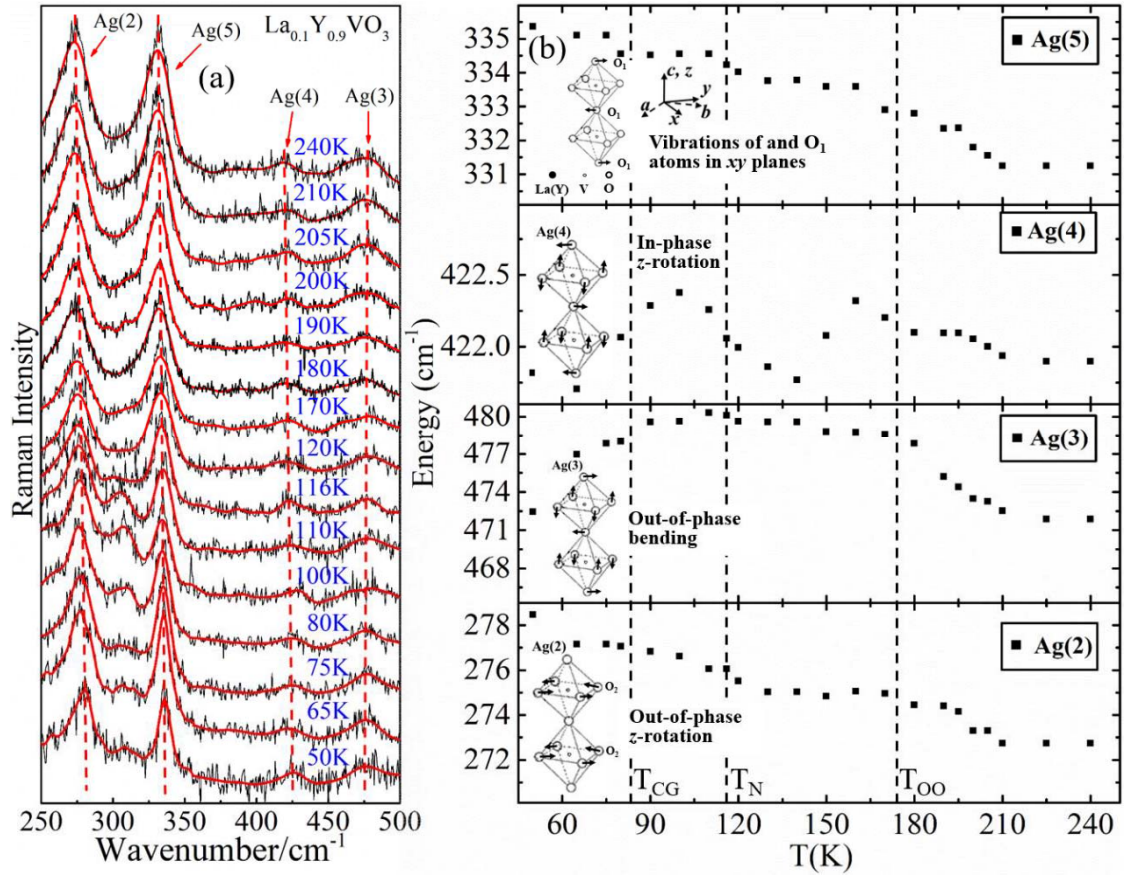

Fig. 6. Temperature dependent Raman spectra (a) and Phonon frequencies (b) for $La_{0.1}Y_{0.9}VO_3$ single crystal in a(bb)a configuration.

To investigate phonons and phonon-related coupling, the Raman spectra were measured, and the frequencies were fitted by a Lorentzian function. Figure. 6 (a) illustrated temperature dependent Raman spectra of $La_{0.1}Y_{0.9}VO_3$ with the *a*(*bb*)*a* configuration, and four dominating peaks associated with the phonon modes Ag(2), Ag(3), Ag(4) and Ag(5) are determined [26]. The Ag(2) and Ag(5) mode corresponds to the basal and apical oxygen vibration in the *ab* plane, respectively, and shift to higher wavenumber with increasing temperature, Fig. 6(b). Although the transitions at $T_{CG}$ and $T_N$ were also obtained at the Ag(5) frequency as the magnetization and heat capacity measurements, they were not obvious for the Ag(2) mode and suggested that the structure transition between Pnma and $P2_1c$ closely related to the apical oxygen vibration in $VO_6$ octahedron.

The Ag(4) vibrational mode is different to the Ag(2) and Ag(5) phonons, which slightly softens from 170K to 140K, and hardens to 100K, finally slowly down to 50 K. Since it is associated with out-of-phase *x*-rotation mode [27], the subtle changes of the phonon mode were due to the competition between the JT coupling and the SE interactions. The Ag(3) phonon mode enhances with decreasing temperature from 240

K to T$_{CG}$, then keeps constant between T$_{CG}$ and T$_N$ around 479 cm$^{-1}$, as temperature reducing continuously the frequency drops to 472.5 cm$^{-1}$ at 50 K. As the mode of the out-of-phase bending [27], the frequency change of the Ag(3) mode should relate the structure change of monoclinic P2$_1$/c from the lattice perturbation.

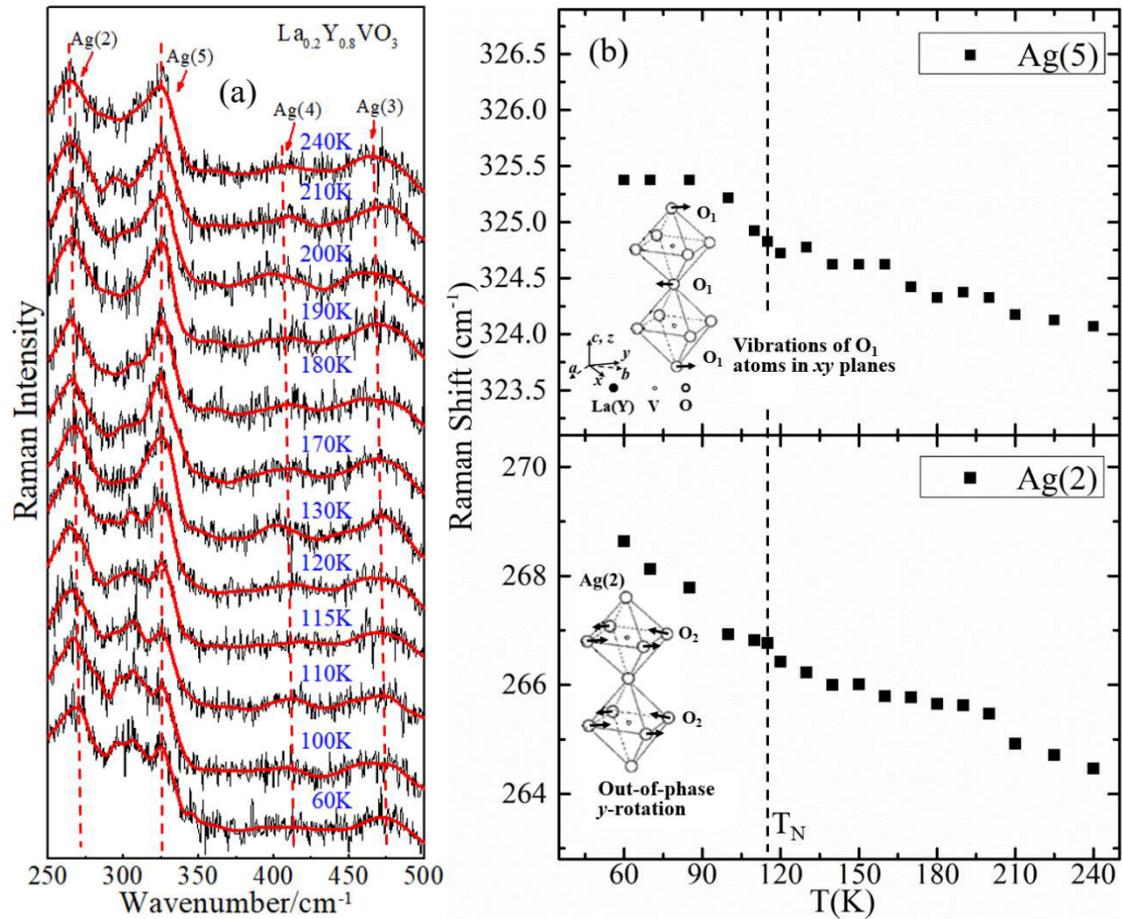

Fig. 7. Temperature dependent Raman spectra (a) and Phonon frequencies (b) for La$_{0.2}$Y$_{0.8}$VO$_3$ single crystal in a(bb)a configuration.

Figure 7 presents the Raman spectra of La$_{0.2}$Y$_{0.8}$VO$_3$, and the Ag(2) and Ag(5) phonon frequencies are hardening with temperature decreasing from 300 K to 50 K. Meanwhile, the frequencies of the Ag(2) and Ag(5) phonons were entirely soften in the whole temperature range compare with that for La$_{0.1}$Y$_{0.9}$VO$_3$ which may be interfered by the lattice distortion from different R-radii. The Ag(3) and Ag(4) phonon frequencies are hard to determine due to the broad FWHM and weak intensity.

## 4. Discussion

The difference in structure, magnetic properties, specific heat, and Raman scattering among La$_x$Y$_{1-x}$VO$_3$ (0≤x≤0.2) were discussed, which are related to the V$^{3+}$ orbital contribution precisely adjusted by La$^{3+}$. With increasing x from 0 to 0.2, the temperature interval between $T_{CG}$ and $T_N$ becomes narrow for $La_xY_{1-x}VO_3$, Fig. 2(a), and the C-AF phase is gradually vanishing. According to the results from Curie-Weiss fitting, Fig. 2(b), we can conclude that the spin-orbital coupling is stronger in $YVO_3$ than that in $La_{0.1}Y_{0.9}VO_3$. This suggests that the substitution of 10% $La^{3+}$ for $Y^{3+}$ ions in $YVO_3$ weakens the spin-orbital coupling just above $T_N$, which decreases the stability of the high temperature C-AF phase. On the other hand, the transition temperatures obtained from the specific heat curves of these samples show the same variation tendency as well as that acquired from the M~T curves. As $x$ increase from 0 to 0.2 in $La_xY_{1-x}VO_3$, the entropy release in $T_{CG}$~$T_N$ reduces and finally disappears at $x=0.2$, which also relates to the vanishing high temperature C-AF phase.

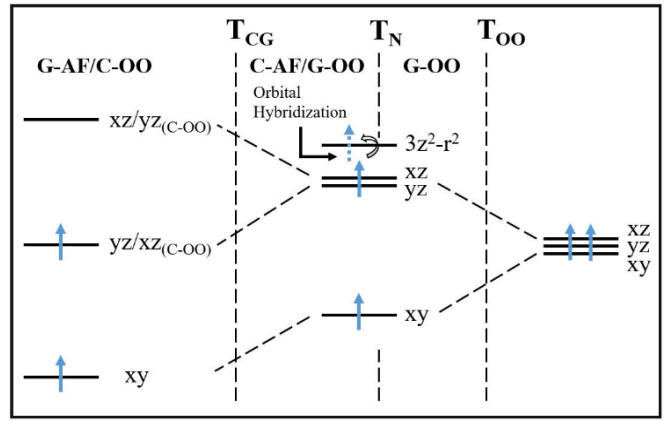

Fig. 8. An schematic view of the $t_{2g}$ orbitals occupancy by the $3d^2$ electron in $YVO_3$ and $La_{0.1}Y_{0.9}VO_3$ in different temperature regimes.

According to the previous research, The G-OO phase is not robust but fluctuations [28] and accompanied with the degenerate of $yz/zx$ orbital. This $yz/zx$ orbital fluctuations may correlate with the entropy release in the whole temperature range of $T_{CG}$~$T_N$ for $La_{0.1}Y_{0.9}VO_3$. Moreover, the slight reductions of both the long bond length along the [001] direction and the octahedral volume on changing from G-OO to C-OO illustrate some $e_g$ orbital ($3z^2-r^2$) occupation owning to a hybridization between $t^2$ and $e_g$ configurations in the G-OO state [29], as shown in Fig.8. This suggests that the anomaly entropy release in the C-AF phase may also be attributed to the lattice perturbation which relates to the random distribution of the $V^{3+}$ ions with higher $e_g$ occupation. Therefore, the existence of the C-AF phase may also benefits to the existence of the $V^{3+}$ ions with $3z^2-r^2$ orbital which increase the SE interaction between NN $V^{3+}$ ions along the c-axis over that in the ab plane ($J_c>J_{ab}$). Furthermore, in the *ab* plane, due to the random distribution of the orbital hybridized $V^{3+}$ ions, the SE interaction in this plane is anisotropy which may introduced the magnetic and lattice anisotropy.

Above all, the shrinking of C-AF phase is related with the weakening of the G-OO phase with increasing $x$ from 0 to 0.1, which indicates the decrease of orbital fluctuation and hybridization. More detailed, the random distribution of the $V^{3+}(3d^2)$ $t_{2g}^2$ and $V^{3+}(3d^2)$ $t_{2g}^1 e_g^1$ ions benefits to the orbital fluctuation. However, in $La_{0.2}Y_{0.8}VO_3$, the C-AF/G-OO phase is hardly observed, while the G-AF/C-OO phase is dominated below $T_N$. In other words, the disappearance of $V^{3+}$ ions with higher $e_g$ occupation in $La_{0.2}Y_{0.8}VO_3$ attributes to the uniform distribution of the $V^{3+}$ ions with $t_{2g}$ occupation, benefiting to the equal SE interaction along the c-axis and in the ab plane ($J_c \approx J_{ab}$), which enhances the C-OO phase. What's more, the strong spin-orbital coupling exits just above $T_N$ may suggest that the short range G-AF/C-OO ordering or spin-orbital entangle exist in $La_{0.2}Y_{0.8}VO_3$.

On the other hand, $J_c$ is closely correlated to the Hund's exchange η which is seriously influenced by the crystal structure [30-33]. In $La_{0.1}Y_{0.9}VO_3$, the monoclinic $P2_1/c$ emerged between $T_{CG}$ and $T_{OO}$ enhances $J_c$ and promotes the stability of the weak G-OO state driven by the SE interaction. In this temperature range, the Ag(3) phonon is hardening from the strength of the out-of-phase bending, which relates to the enhancement of $J_c$. The C-OO phase is induced by JT coupling, while the G-OO state is favored by the SE interaction. The temperature dependent competition between the SE interaction and JT coupling relates to the abnormal behavior of Ag(4) phonon as shown in Fig.6 (b), which also suggests that SE interaction dominants in the G-OO existed temperature range. From $La_{0.2}Y_{0.8}VO_3$, the broad FWHM and weak intensity of Ag(4) phonon suggest the diminish of the competition mechanism and the lattice perturbation.

Except affecting the NN $V^{3+}$ ions, the tilting of $VO_6$ in the distorted lattice of $La_{1-x}Y_xVO_3$ also leads to non-collinear magnetic moments in an antiferromagnet. Such tilting introduces staggered V-O bonds, which staggers the easy axis of single ion magnetic anisotropy [39]. On the other hand, owning to the $VO_6$ octahedra tilting the oxygen ion between the two nearest-neighbor (NN) V ions are not at an inversion center, which gives rise to a DM interaction [34-36]. The microscopic form of DM interaction is $D \cdot (S_A \times S_B)$, where $S_A$ and $S_B$ are two neighboring spins and D is the Moriya one-bond vector [35, 36]. The single-ion magnetic anisotropy and DM interaction determine the canting angle, and the anomalous diamagnetism is suggested [37]. The net magnetization along the $a$-axis was contributed by the AFM interaction, DM interaction

and single ion anisotropy [25]. The net energy derived from the expression is

$$\frac{E}{N} = J\mathbf{S}_A \cdot \mathbf{S}_B + \mathbf{D} \cdot (\mathbf{S}_A \times \mathbf{S}_B) - 2AS_Z^2 + \frac{E_f}{N} \quad (2)$$

Where $J$ is the AFM coupling constant, $\mathbf{D}$ is the Moriya one-bond vector, $A$ is the single-ion anisotropy constant, $S_Z$ is the component of spin along the local easy axis and $E_f$ is the NN ferromagnetic interaction which does not relate to the canting angle. By introducing the spin canting angle Eq. (2) can be written as

$$\frac{E}{N} = -J|<S>|^2\cos(2\theta) + D|<S>|^2\sin(2\theta) - 2AS^2\cos^2(\theta - \gamma) + \frac{E_f}{N} \quad (3)$$

where $\theta$ is the canting angle with respect to the easy axis, $\gamma$ is the canting angle owning to single ion anisotropy. In $La_xY_{1-x}VO_3$ (0≤$x$≤0.2) system, ($\theta$-$\gamma$) << 1. By minimizing the energy, sine and cosine functions can be simplified by the first term of their Taylor expansion.

$$\theta = \frac{S^2\xi\gamma - \gamma_D|<S>|^2}{|<S>|^2 + S^2\xi} \quad (4)$$

Where $\xi=A/J$, $\gamma_D=D/2J$ and S=1. According to the mean field model the net magnetic moment of the AFM can be written as

$$\begin{cases} M_{net} = 2Ng\mu_B|<S>|\sin(\theta) \approx 2Ng\mu_B|<S>|\theta \\ |<S>| = \frac{g\mu_B S(S+1)H}{3k_B T} \end{cases} \quad (5)$$

On the basis of Eq. (5), the temperature dependence of magnetizations can be calculated, Fig. 2 (a). Based on the fitting results, the parameters A, D and J as shown in Table. 2 can describe all the microscopic features of the temperature-dependent net magnetic moment in the interval of $T_{CG}$<T<$T_N$.

Table 2. The fit parameters of $La_xY_{1-x}VO_3$ (0≤$x$≤0.2)

|   | YVO$_3$ | La$_{0.1}$Y$_{0.9}$VO$_3$ | La$_{0.2}$Y$_{0.8}$VO$_3$ |
|---|---|---|---|
| A | 435.0 K | 513.3 K | 828.3 K |
| D | 14.2 K | 7.0 K | 3.3 K |
| γ | 6.93×10$^{-3}$ | 6.6×10$^{-3}$ | 6.0×10$^{-3}$ |

The estimated values of single ion anisotropy and DM constant of the present samples are similar to YVO$_3$ single crystal [25]. The values of A increases, while D drops with rising $x$, Table 2, thus, the increasing La$^{3+}$ content in $La_xY_{1-x}VO_3$ weakens the DM interaction and enhance the single ion anisotropy to destabilize the anomalous diamagnetism. Meanwhile, Fig. 4(a) illustrate the field-dependence of the diamagnetism which is attributed to the decrease of the DM interaction and the enhancement of the single ion anisotropy. For $La_{0.2}Y_{0.8}VO_3$, the field-dependence of M-T curves present the gradually change from ferromagnetic to antiferromagnetic with the decreasing canting angle and the weak DM interaction is suggested.

## 5. Conclusion

The mechanism of the anomalous magnetism and evolution in low La-doped $La_xY_{1-x}VO_3$ ($0 \leq x \leq 0.2$) single crystals have been investigated using XRD, magnetization, specific heat and Raman scattering measurements. It is found that large NN exchange interaction stabilize the fluctuant G-OO and favor the exotic C-AF, which relates to large entropy release. With increasing the $La^{3+}$ content, the orbital fluctuation and hybridization were weakened to induce C-OO and labilized G-OO, meanwhile, the anisotropic NN exchange interaction in the *ab* plane could become smaller. What's more, short range spin-orbital correlation may exist in $La_{0.2}Y_{0.8}VO_3$ at temperatures just above $T_N$. Furthermore, through a theoretical model, we find that the increasing $La^{3+}$ content in $La_xY_{1-x}VO_3$ weakens the DM interaction and enhance the single ion anisotropy to destabilize the anomalous diamagnetism. Finally, for $La_{0.2}Y_{0.8}VO_3$ the magnetic field weakens the DM interaction and interfere the magnetization from ferromagnet to antiferromagnet by decreasing the canting angle.

## Acknowledgements

J.M., G.T.L. thank the financial support from the National Science Foundation of China (No. 11774223, U2032213, U1732154, and 12004243). G.T.L. thanks the project funded by China Postdoctoral Science Foundation (Grant No. 2019M661474). J.M. acknowledges additional support from SJTU-KTH Collaborative Research and Development Seed Grants in 2020. X.Y. is supported by Key Projects for Research & Development of China (Grant No. 2016YFA0300403).